\documentclass[9pt,twocolumn,twoside]{osajnl}

\journal{ol} 

\setboolean{shortarticle}{true}

\title{Unidirectional vortex waveguides and multistable vortex pairs in polariton condensates}

\author[1]{Xinghui Gao}
\author[2]{Wei Hu}
\author[3,4]{Stefan Schumacher}
\author[3,*]{Xuekai Ma}

\affil[1]{School of Electronics and Communication Engineering, Guangzhou University, Guangzhou 510006, P. R. China}

\affil[2]{Guangdong Provincial Key Laboratory of Nanophotonic Functional Materials and Devices, South China Normal University, Guangzhou 510631, P. R. China}

\affil[3]{Department of Physics and Center for Optoelectronics and Photonics Paderborn (CeOPP), Universit\"{a}t Paderborn, Warburger Strasse 100, 33098 Paderborn, Germany}
\affil[4]{Wyant College of Optical Sciences, University of Arizona, Tucson, AZ 85721, USA}

\affil[*]{Corresponding author: xuekai.ma@gmail.com}

\begin{abstract}
Vortices carrying quantized topological charges have potential application in information processing. In this work, we investigate vortex carriers and waveguides in microcavity polariton condensates, nonresonantly excited by a homogeneous pump with intensity grooves. An intensity groove with a ring shape in the pump gives rise to dark-ring states of the condensate with a $\pi$-phase jump, akin to dark solitons. The dark-ring states can be destroyed by a stronger density of the surrounding condensate and reduce into vortex-antivortex pairs. Multiple vortex-pair states are found to be stable in the same dark ring of the pump. When the pump ring is broader, higher-order dark states with multiple $\pi$-phase jumps can be obtained and interestingly they can be used to construct vortex waveguides. If a single vortex is imprinted in such waveguides, it can travel in a particular direction, showing one-way transportation. In other words, an imprinted vortex with a certain charge in a specifically designed higher-order dark state is only allowed to propagate unidirectionally. 

\end{abstract}
\setboolean{displaycopyright}{true}

\begin{document}

\maketitle
Transmission of quantized information has attracted significant attention, since a quantized signal is less distorted compared with a continuous signal during the propagation. As such kind of information carriers, vortices have been widely reported in different physical systems. A vortex has a quantized winding number, known as topological charge, representing the rotation of the particles enclosing a phase defect. The propagation of vortices has been intensively studied in nonlinear optics where a vortex soliton can propagate over a long distance without decay in either intensity or phase~\cite{kivshar2003optical}.

Among the different systems that exhibit vortex formation, vortices in exciton-polariton (polariton for short) systems \cite{lagoudakis2008quantized,lagoudakis2009observation,sanvitto2010persistent,roumpos2011single,nardin2011hydrodynamic,PhysRevLett.113.200404,PhysRevLett.121.225302,ma2020realization} have recently attracted considerable attention. Polaritons are light-matter hybrid quasiparticles composed of strongly coupled excitons and photons in semiconductor microcavities. Polariton vortices can be created both resonantly by coherent imprinting and nonresonantly due to the spontaneous relaxation and the formation of macroscopic coherence (known as polariton condensation)~\cite{deng2002condensation,kasprzak2006bose}. It is found that in nonequilibrium polariton condensates, one-dimensional dark solitons are unstable in an otherwise homogeneous background~\cite{smirnov2014dynamics,xue2014creation}, but a modulated background can stably confine them~\cite{PhysRevLett.118.157401}. Dark soliton stripes in two dimensions can evolve into vortex-antivortex pairs~\cite{amo2011polariton,PhysRevB.83.144513,PhysRevLett.107.245301,smirnov2014dynamics} due to the transverse instability and the interaction of the polariton vortices typically results in complex in-plane motion~\cite{fraser2009vortex,PhysRevB.81.235302,PhysRevLett.107.036401,sanvitto2011all,nardin2011hydrodynamic,dominici2015vortex,PhysRevB.91.085413,dominici2018interactions,PhysRevB.102.045309}. Such vortices can also be confined and manipulated in two-dimensional optical lattices~\cite{PhysRevLett.118.157401}. However, efficient control or vortex transportation to a certain destination has not been addressed in polariton condensates, which is promising for application and may be achieved in optically induced channels that are able to transport phase defects~\cite{PhysRevB.96.205406}.

\begin{figure}[t]
\centering
{\includegraphics[width=1\linewidth]{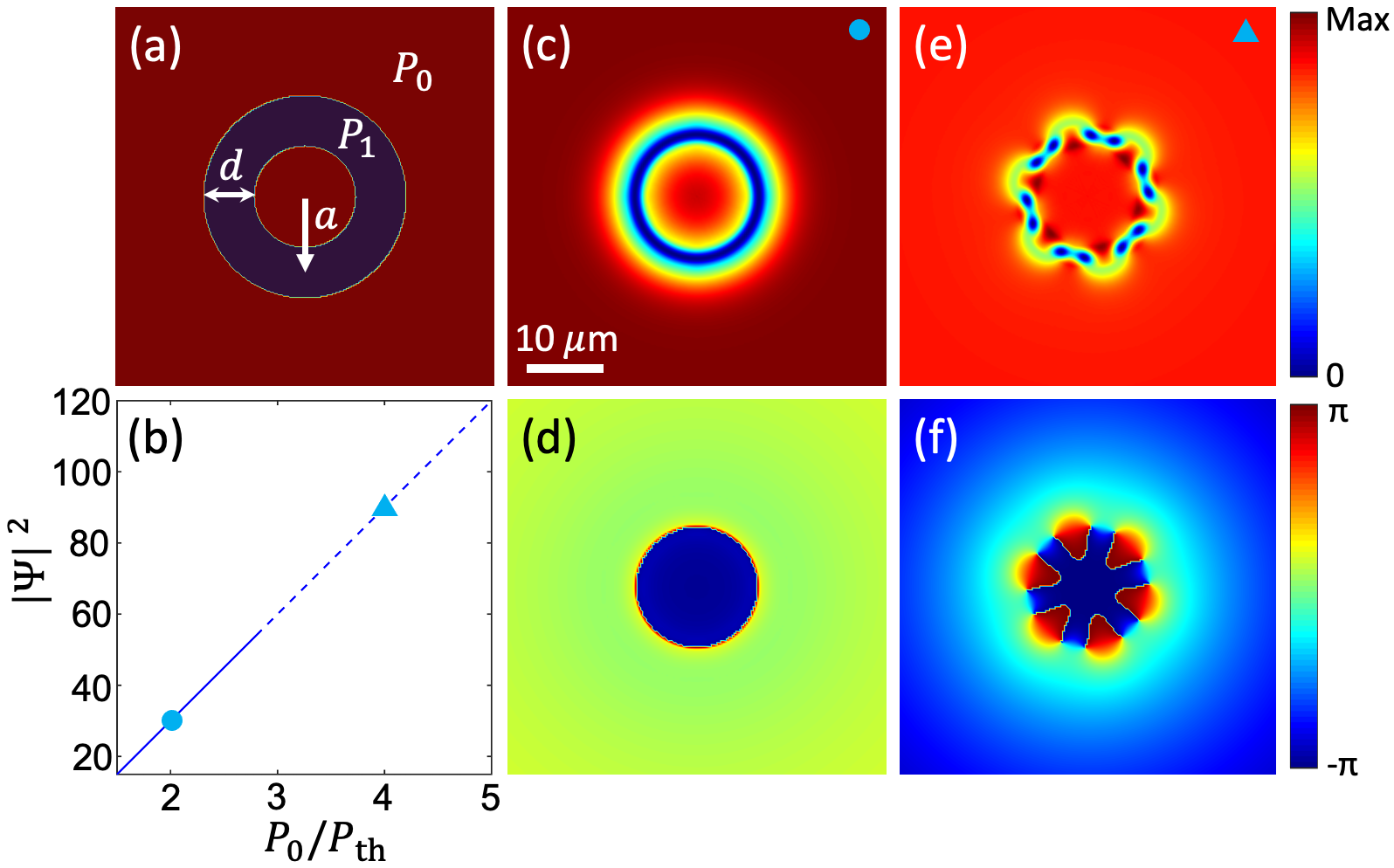}}
\caption{\textbf{Excitation scheme and solutions.} (a) Sketch of a homogeneous pump with a ring-shaped intensity gap (dark ring). $P_0$ is the background pump intensity. $P_1$, $d$, and $a$ represent respectively the intensity, width, and radius of the lower-intensity ring in the pump. (b) Dependence of the peak background density $|\Psi|^2$ (in $\mu$m$^{-2}$) of the polariton condensate on the background pump intensity $P_0$ at $P_1=0$, $a=8$ $\mu$m, and $d=3$ $\mu$m. The solid line represents stable dark-ring states, while the dashed line represents the region where the dark ring is unstable but evolves into stable vortex pairs. (c,e) Density and (d,f) phase profiles of the dark-ring and vortex-pair states, respectively, corresponding to the states marked in (b).}
\label{fig:1}
\end{figure}

In this Letter we excite polariton condensates by using a dark-ring pump, i.e., a homogeneous pump with a ring-shaped intensity groove as shown in Fig. \ref{fig:1}(a). This lower intensity ring in the pump can support stable dark-ring states with a $\pi$-phase jump. As the pumping intensity increases above a threshold value, the dark-ring state becomes unstable, known as snake instability~\cite{PhysRevLett.123.215301,Claude:20}, and evolves into several persistent vortex-antivortex pairs. We find that the number of the vortex pairs induced is related to the radius of the dark ring in the pump. Moreover, in some cases, more than one vortex-pair state can be stabilized in the same pump ring, forming multistable vortex-pair states. The stability of each vortex-pair state can be increased by the surrounding reservoir density via the polariton-reservoir interaction. If the pump ring is broader, higher-order vortex-pair states with multiple vortex-pair layers can be created and in each layer the number of the vortex pairs is identical, which tends to weaken the snake instability. When the snake instability is prevented, higher-order dark-ring states, which can be used to design vortex waveguides to unidirectionally transport vortices with specific topological charges, can be excited.  

To mimic the dynamics of microcavity polariton condensates under nonresonant excitation, we use a driven-dissipative Gross-Pitaevskii model coupled to a rate equation describing the density of the incoherent reservoir ~\cite{wouters2007excitations}:
\begin{equation}\label{e1}
\begin{aligned}
i\hbar\frac{\partial\Psi(\mathbf{r},t)}{\partial t}&=\left[-\frac{\hbar^2}{2m_{\text{eff}}}\nabla_\bot^2-i\hbar\frac{\gamma_\text{c}}{2}+g_\text{c}|\Psi(\mathbf{r},t)|^2 \right.\\
&+\left.\left(g_\text{r}+i\hbar\frac{R}{2}\right)n(\mathbf{r},t)\right]\Psi(\mathbf{r},t),
\end{aligned}
\end{equation}
\begin{equation}\label{e2}
\frac{\partial n(\mathbf{r},t)}{\partial t}=\left[-\gamma_r-R|\Psi(\mathbf{r},t)|^2\right]n(\mathbf{r})+P(\mathbf{r},t)\,.
\end{equation}
Here, $\Psi(\mathbf{r},t)$ is the wavefunction of the polariton condensate and $n(\mathbf{r},t)$ is the density of the reservoir. The effective mass of the condensate is denoted by $m_{\text{eff}}=10^{-4}m_\textup{e}$ ($m_\textup{e}$ is the free electron mass). $\gamma_\text{c}=0.2$ ps$^{-1}$ and $\gamma_\text{r}=0.3$ ps$^{-1}$ represent the decay rate of the condensate and the reservoir, respectively. The condensate is replenished by the excitons from the reservoir with a condensation rate $R=0.01$ ps$^{-1}$. $g_\text{c}=6$ $\mu$eV $\mu$m$^2$ describes the polariton-polariton interaction and $g_\text{r}=2g_\text{c}$ gives the polariton-reservoir interaction. These parameters are typical for GaAs-based semiconductor microcavities~\cite{roumpos2011single}. The nonresonant excitation $P(\mathbf{r},t)$ is a continuous-wave pump composed of a homogeneous background and a ring-shaped intensity gap:
\begin{equation}\label{pump}
P(\mathbf{r})=
\begin{cases}
P_1 & \ \ \ \ \ \ a-d/2<\mathbf{r}<a+d/2 \\
P_0 & \ \ \ \ \ \ \textup{elsewhere}
\end{cases}
\end{equation}
Here, $P_1$ and $P_0$ represent the intensity in and out of the dark ring, respectively, with $P_1<P_0$. The radius and width of the dark ring are governed by $a$ and $d$, respectively, as can be seen in Fig. \ref{fig:1}(a). Under homogeneous excitation, the condensation threshold is given by $P_\textup{th}=\gamma_\textup{c}\gamma_\textup{r}/R$. Here, the coupled equations are solved by using the fourth-order Runge-Kutta method.

\begin{figure}[t]
\centering
{\includegraphics[width=1\linewidth]{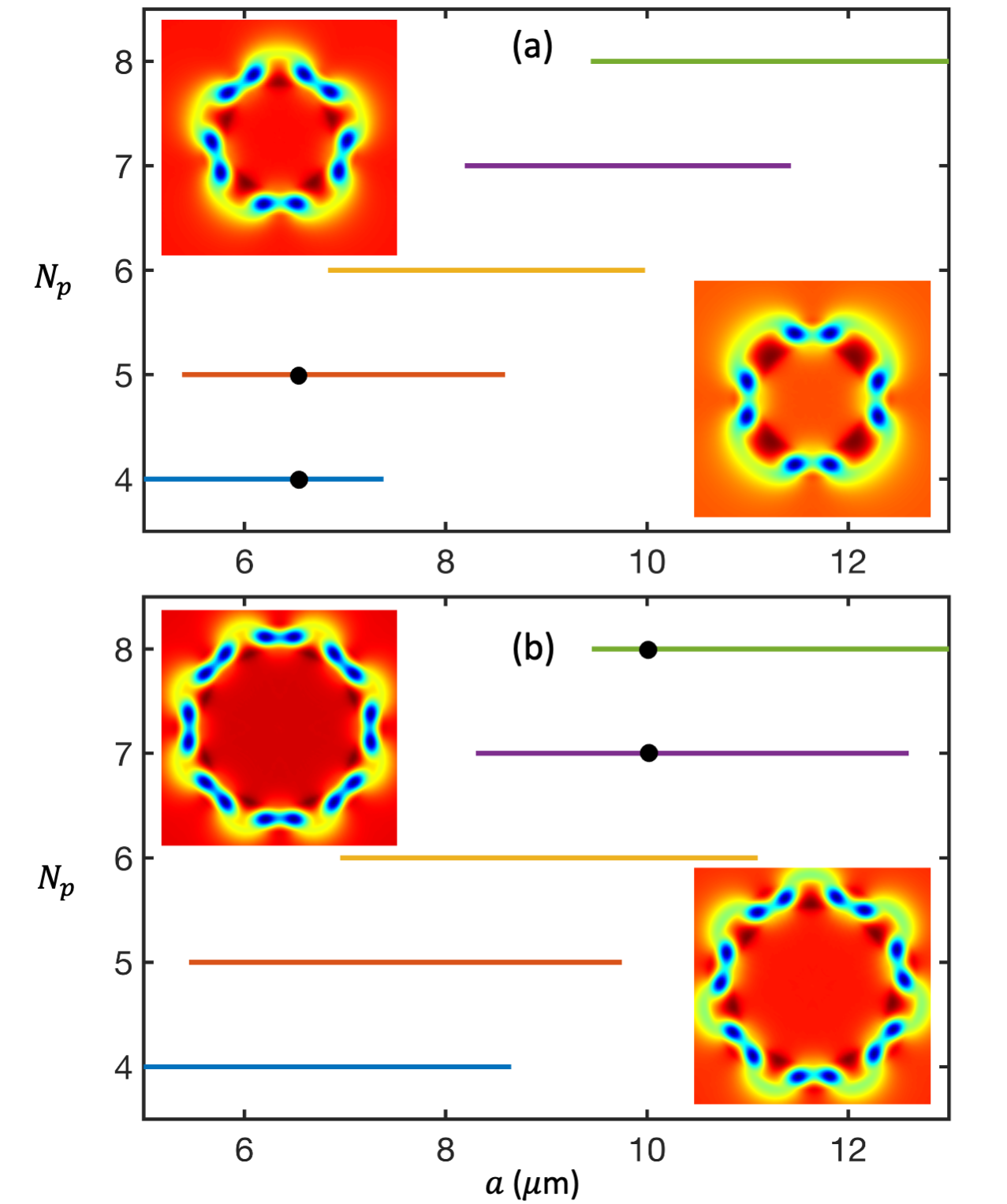}}
\caption{\textbf{Multistable vortex-pair states.} Dependence of the number of the vortex pairs $N_p$ in the vortex-pair states on the radius $a$ of the dark pump ring with intensity (a) $P_1=0$ and (b) $P_1=0.2P_\textup{th}$. The colored lines represent the regions of the stable solutions. The inserts correspond to the states marked by the black dots, respectively. Here, $d=3$ $\mu$m and $P_0=4P_\textup{th}$.}
\label{fig:2}
\end{figure}

A lower intensity ring with $a=8$ $\mu$m and $d=3$ $\mu$m in the pump leads to the formation of a dark-ring state with a $\pi$-phase jump [Fig. \ref{fig:1}(c,d)]. Similar dark states can also be supported by ring-shaped defects~\cite{jheng2019ring}. This state is stable as long as the pump intensity $P_0$ is not far above the condensation threshold [Fig. \ref{fig:1}(b)]. Note that to create such dark-ring states and remove the background instability~\cite{PhysRevB.91.085413}, $P_0$ should not be chosen too close to the condensation threshold. For the parameters mentioned above, the dark-ring state becomes unstable when $P_0>2.8P_\textup{th}$ as presented in Fig. \ref{fig:1}(b). This critical point changes for different values of the parameters. The instability results in the decay of the dark-ring state and consequently the formation of vortex-antivortex pairs [Fig. \ref{fig:1}(e,f)]~\cite{PhysRevLett.123.215301}. The vortex pairs are also stable steady-states without motion or further decay and their specific positions in the ring are random depending on the surrounding environment. The stability of all the states are confirmed by adding complex (amplitude and phase) broadband noise and letting them evolve over long time (thousands of polariton lifetimes $1/\gamma_\textup{c}$). The feedback of the solutions in Fig. \ref{fig:1}(c,e) according to \eqref{e2} leads to different reservoir distributions.

Figure \ref{fig:2} shows the dependence of the stability of different vortex-pair states on the radius of the dark pump ring. Different vortex-pair states refer to the states having different numbers of vortex pairs $N_p$. If the intensity of the dark ring in the pump vanishes completely with $P_1=0$, one can see in Fig. \ref{fig:2}(a) that more than one vortex-pair state can stably survive against noise. For example, both vortex-pair states with $N_p=4$ and $N_p=5$ pairs [see the inserts in Fig. \ref{fig:2}(a)] are stable at $a=6.5$ $\mu$m. At some points, $a=7$ $\mu$m for instance, even three vortex-pair states are stable. Which vortex-pair state can be selected depends on the initial noise and the fluctuation during the time evolution.

Interestingly, the stability of each vortex-pair state against the radius of the pump ring $a$ can be enhanced by the polariton-reservoir interaction. As can be seen in Fig. \ref{fig:2}(b), where the intensity in the pump ring is increased to $P_1=0.2P_\textup{th}$, the reservoir density in the pump ring offers additional interaction, further stabilizing the vortex pairs. If $P_1$ increases further but is still below the threshold, the vortex-pair states become more stable. The vortex-pair states carrying quantized topological charges can be regarded as information carriers or containers, with the possibility to switch between them. To this end, one can increase or decrease back and forth the radius of the pump ring during the excitation to let the current state jump to the desired states.

\begin{figure} [t]
\centering
{\includegraphics[width=1\linewidth]{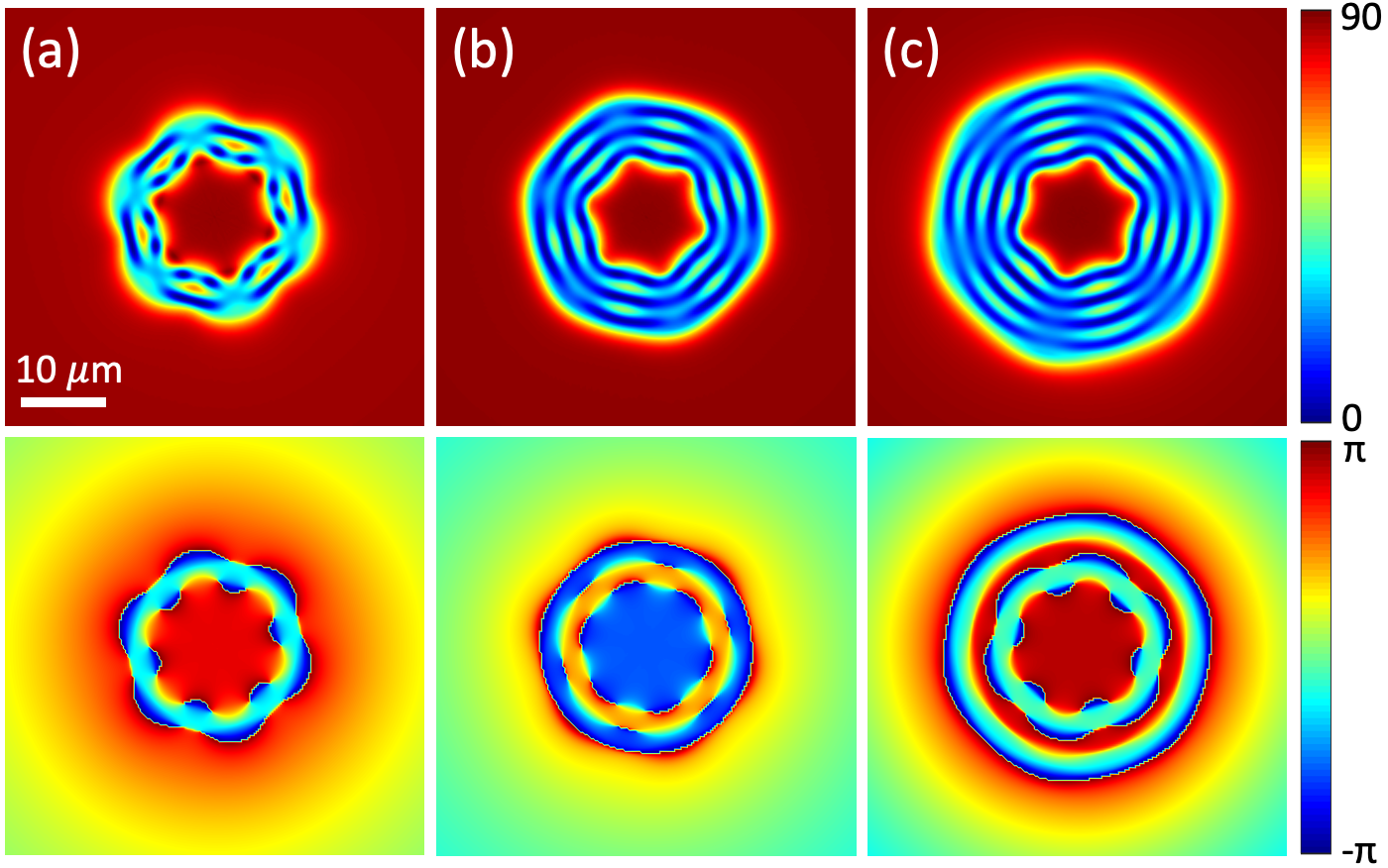}}
\caption{\textbf{Higher-order vortex-pair states.} Density (upper row) and phase (lower row) profiles of vortex-pair states formed in different pump rings with (a) $a=9$ $\mu$m and $d=5$ $\mu$m, (b) $a=10$ $\mu$m and $d=7$ $\mu$m, and (c) $a=11.5$ $\mu$m and $d=10$ $\mu$m. For all the cases, $P_1=0$ and $P_0=4P_\textup{th}$.}
\label{fig:3}
\end{figure}

Besides the radius of the dark pump ring, its width also affects the states being trapped. Examples can be found in Fig. \ref{fig:3} where higher-order vortex-pair states with multiple layers of vortex-pair rings are obtained and the number of the layers increases with the width of the pump ring. In Fig. \ref{fig:3}, the width of the pump ring is increased by expanding the size of the outer edge of the ring and keeping the position of its inner edge fixed, in order to always obtain the vortex-pair states with $N_p=6$ (using the same initial condition as in Fig. \ref{fig:1}) for the comparison. Since in each layer the number of the vortex pairs are identical, the vortex pairs in the most outer ring layer, such as in Fig. \ref{fig:3}(c), become strongly stretched. The deformed outer vortex pairs react to the inner ones, followed by the deformation of the inner vortex pairs, which tends to prevent the snake instability. As a consequence, the vortex pairs in Fig. \ref{fig:3}(c) are hard to be recognized in contrast to the ones in Fig. \ref{fig:3}(a).

\begin{figure}[t]
\centering
{\includegraphics[width=1\linewidth]{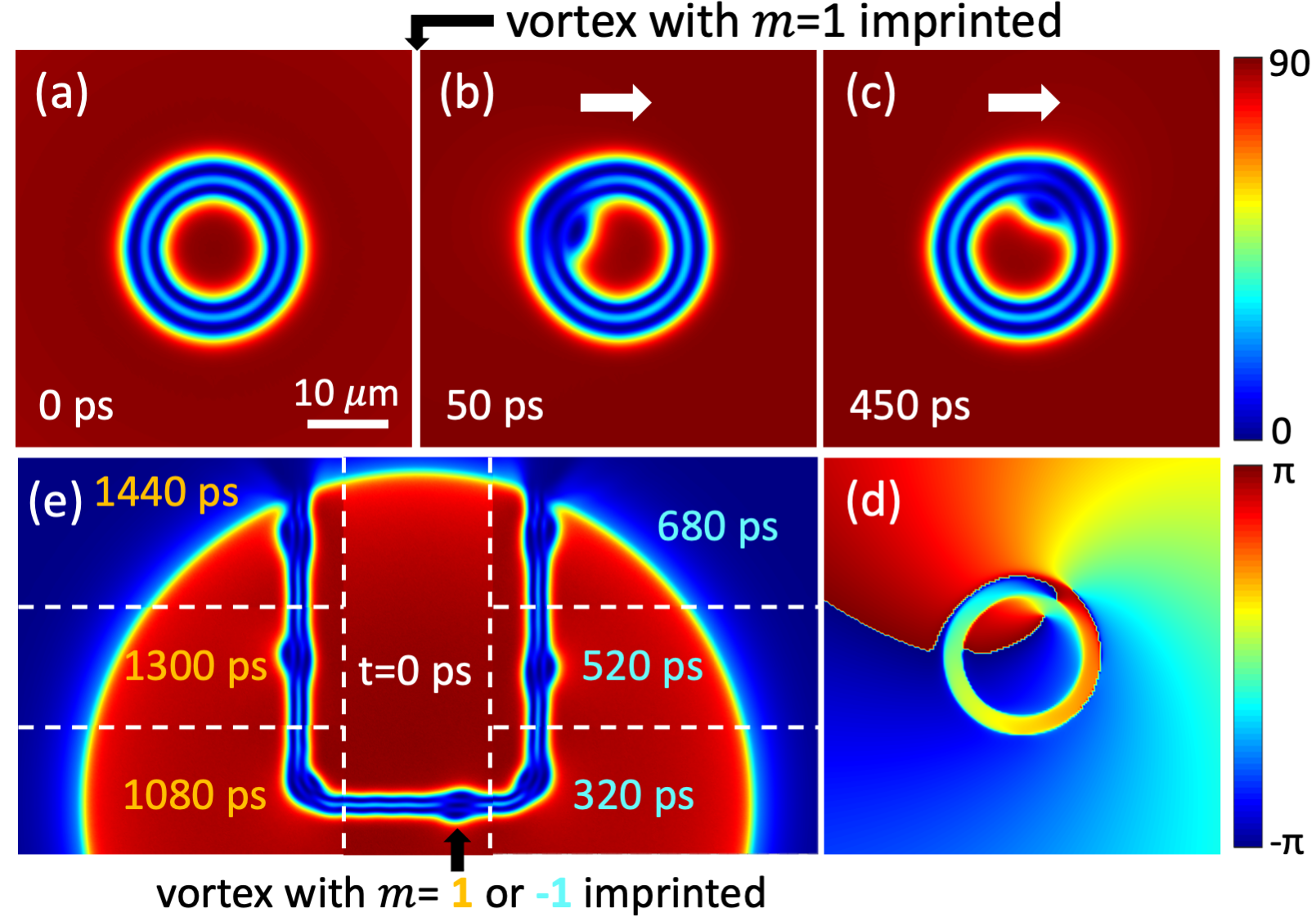}}
\caption{\textbf{Vortex waveguides.} (a-c) Density profiles of a higher-order dark-ring state carrying a vortex with $m=1$, which is imprinted to the center of the state in (a), in (b,c) at different time. The white arrows indicate the traveling direction of the vortex. (d) Phase profile of the state in (c). For (a-c), $a=8.5$ $\mu$m, $d=4$ $\mu$m, $P_0=4P_\textup{th}$, and $P_1=0$. (e) Density profile of a U-shaped higher-order dark state excited by a tailored pump with an intensity groove that has the same U-shape profile with $d=4$ $\mu$m, $P_0=4P_\textup{th}$, and $P_1=0$, and a broad intensity filter with the diameter of 170 $\mu$m. The time in cyan indicate the positions of the moving vortex of $m=-1$, while the time in orange indicate the positions of the moving vortex of $m=1$. The vortex in each simulation is imprinted at the same position marked by the black arrow.}
\label{fig:4}
\end{figure}

The snake instability can also be prevented when the state transitions from the one in Fig. \ref{fig:1}(e,f) to that in Fig. \ref{fig:3}(a) by gradually increasing the radius of the pump ring [see Fig. \ref{fig:4}(a)]. In this scenario, a higher-order dark-ring state that carries two $\pi$-phase jumps forms. The higher-order dark-ring states are stable at weaker pump intensity and become unstable at higher pump intensity. Surprisingly, this dark state can support traveling of the vortices imprinted. For example, we take the solution in Fig. \ref{fig:4}(a) as the initial condition and immediately imprint a vortex with $m=1$ by multiplying a phase singularity in the ring region. One can see that the vortex starts to travel clockwise in the ring, which is different from that caused by the Magnus force~\cite{wingenbach2021dynamics}, because in this case the background density is almost homogeneous without a density gradient, or the Josephson vortices formed between two condensate rings~\cite{PhysRevB.104.165305} in which the motion of the vortex depends on the coupling strength of the rings. If a vortex with topological charge $m=-1$ is imprinted at the beginning, it circulates counter-clockwise. Therefore, the higher-order dark-ring states can be treated as vorticity screeners or vortex waveguides. Note that the dark-ring state in Fig. \ref{fig:1}(c) cannot be used to transport vortices as in the case of Fig. \ref{fig:3}(b,c), because for the former the vortex core is much larger than the width of the dark-ring state, so that the vortices are difficult to be captured and later carried by the dark ring in the condensate. 

Dark states similar to the ones in Figs. \ref{fig:1}(c) and \ref{fig:4}(a) can also be excited by a homogeneous pump with a linear instead of a ring-shaped intensity groove. However, a vortex imprinted in a straight dark state [cf. the middle part in Fig. \ref{fig:4}(e)] is spatially pinned, even though the flat background condensate has a finite extent, without moving in virtue of the symmetric configuration and the higher background density. To successfully transport a phase defect to a certain position, we design a U-shaped intensity groove in a broad pump with a finite size. The finite size of the broad pump is to remove the influence of the boundaries in the simulation. As expected, under such excitation, the same U-shaped dark state in the condensate is created as shown in Fig. \ref{fig:4}(e). After that, we imprint a vortex slightly to the right side of the pump center [indicated by the black arrow in Fig. \ref{fig:4}(e)] and record the motion of the vortex over time. If the vortex takes the topological charge $m=-1$, it circulates counter-clockwise in dark-ring states as analyzed above, so that the vortex in Fig. \ref{fig:4}(e) begins to move towards the right corner of the U-shaped waveguide. After traveling through the corner, it propagates further and finally escapes from the pumping region. The total traveling time is around 700 ps. If a vortex with $m=1$, which circulates clockwise in a dark-ring state as illustrated in Fig. \ref{fig:4}(b,c), is initially imprinted at the same position as the black arrow indicates in Fig. \ref{fig:4}(e), the right corner is forbidden for it, because it cannot propagate counter-clockwise, and finally it travels towards the left corner of the U-shaped waveguide to fulfil the clockwise propagation. But as the initial position of the vortex imprinted is far away from the left corner and it also needs to overcome the outgoing flow of the condensate, induced by the finite pump profile, at the beginning of its travel, it takes more than 1000 ps to reach the left corner. After passing the corner, only less than 400 ps is spent to finish the rest of the travel until escaping the pumping area, which is similar to the speed of the $m=-1$ charged vortex after passing the right corner. Therefore, the unidirectional transport of the specifically charged vortices has been achieved in a U-shape dark state. Note that the transportation dynamics of the vortices shown in Fig. \ref{fig:4} hold when $P_0$ is slightly larger than zero in the case that the higher-order dark ring states are stable, or the vortex is imprinted at other positions, the middle of the U-shaped waveguide for instance. It is also worth noting that the unidirectional propagation of the vortices as well as the multistable vortex pairs remain largely unaffected if a reasonable disorder potential~\cite{PhysRevLett.118.157401} is applied.

To conclude, we have studied the dark states of polariton condensates induced nonresonantly by intensity grooves in the otherwise spatially homogeneous pump. A narrower intensity groove in the pump can support a dark state with a $\pi$-phase jump which becomes unstable as the background pumping intensity increases and evolves into stable vortex-antivortex pairs. The intensity grooves in the pump can be regarded as phase defect containers that can carry multistable vortex pairs. A broader intensity groove in the pump, however, can lead to the formation of higher-order dark states that can be used to unidirectionally transport quantized topological charges, giving rise to the design of more complex vortex waveguides to explore their potential application in information transmission and screening in all-optical circuits.

\textbf{Funding.} National Natural Science Foundation of China (11804064); Deutsche Forschungsgemeinschaft (DFG) (No. 231447078, 270619725); Guangdong Basic and Applied Basic Research Foundation, China (No. 2020A1515010930); Paderborn Center for Parallel Computing, PC$^2$.

\textbf{Data availability.} Data underlying the results presented in this paper are not publicly available at this time but may be obtained from the authors upon reasonable request.

\textbf{Disclosures.} The authors declare no conflicts of interest.

\providecommand{\noopsort}[1]{}\providecommand{\singleletter}[1]{#1}%


\end{document}